\documentclass[aps,prc,showpacs,twocolumn,superscriptaddress]{revtex4}
\usepackage{multirow}
\usepackage{graphicx}
\usepackage{amsmath}
\usepackage{mathrsfs}

\def\version{version}

\newcommand{ \be }{\begin{equation}}
\newcommand{ \ee }{\end{equation}}
\newcommand{ \bea }{\begin{eqnarray}}
\newcommand{ \eea }{\end{eqnarray}}

\def\sNN{\mbox{$\sqrt{s_{_{NN}}}$}}

\def \auau  {$\mathrm{Au}+\mathrm{Au}$ }
\def \cucu  {$\mathrm{Cu}+\mathrm{Cu}$ }

\usepackage{color}

\begin{document}
\title{
\begin{flushright}
{
\small \sl \version 1.4\\
}
\end{flushright}
Study of triangular flow $v_3$ in Au+Au and Cu+Cu collisions with a multiphase transport model}

\author{Kai Xiao}\affiliation{Institute of Particle Physics, Central China Normal University, Wuhan, Hubei, 430079, China}
\affiliation{The Key Laboratory of Quark and Lepton Physics (Central China Normal University), Ministry of Education, Wuhan, Hubei, 430079, China}
\author{Na Li}\email{nli@mail.hust.edu.cn}\affiliation{Department of Physics, Huazhong University of Science and Technology, Wuhan 430074, China}
\author{Shusu Shi}\email{sss@iopp.ccnu.edu.cn}\affiliation{Institute of Particle Physics, Central China Normal University, Wuhan, Hubei, 430079, China}
\affiliation{The Key Laboratory of Quark and Lepton Physics (Central China Normal University), Ministry of Education, Wuhan, Hubei, 430079, China}
\author{Feng Liu}\affiliation{Institute of Particle Physics, Central China Normal University, Wuhan, Hubei, 430079, China}
\affiliation{The Key Laboratory of Quark and Lepton Physics (Central China Normal University), Ministry of Education, Wuhan, Hubei, 430079, China}

\date{\today}

\begin{abstract}

We studied the relation between the initial geometry anisotropy and the anisotropic flow in a multiphase transport model (AMPT) for both Au+Au and Cu+Cu collisions at \sNN=200 GeV. It is found that unlike the elliptic flow $v_2$, little centrality dependence of the triangular flow $v_3$ is observed. After removing the initial geometry effect, $v_3/\varepsilon_3$ increases with the transverse particle density, which is similar to $v_2/\varepsilon_2$. The transverse momentum ($p_T$) dependence of $v_3$ from identified particles is qualitatively similar to the $p_T$ dependence of $v_2$.

\end{abstract}
\pacs{25.75.Ld, 25.75.Dw}

\maketitle
\clearpage
\section{Introduction}
\label{sect_intro}

A novel state of matter called quark-gluon plasma (QGP) composed by deconfined partons is believed to be created experimentally in heavy ion collisions at RHIC~\cite{QGP}. The discovery of large elliptic flow indicates that the partonic collectivity is built up during the collisions, and the number-of-quark scaling suggests that the partonic degrees of freedom are active~\cite{nqscaling}.

The anisotropic flow is usually described by a Fourier decomposition of the azimuthal distribution with respect to the reaction plane~\cite{ArtPRC}. The second harmonic coefficient, $v_2$, so called elliptic flow, has the biggest magnitude at high energy collisions~\cite{largeflow}. It is believed that the observed anisotropy in the momentum space is caused by the anisotropy in the coordinate space in the initial condition. Lots of attention has been put on the relation between $v_2$ and spatial eccentricity to see the hydrodynamics behavior of the created system~\cite{v2ecc, PRCrun4, cucu_STAR}.

Recent studies show that the event-by-event fluctuation of the initial geometry~\cite{v3ecc} may play an important role in the study of collective flow. The triangular shape in the initial geometry will be transferred to the momentum space as the system expands, and finally leads to the none zero value of the third harmonic coefficient, $v_3$. It is found that the triangular flow $v_3$ is responsible for the ridge and shoulder structures and the broad away-side of two-particle azimuthal correlation~\cite{v3ridge}. Besides, it is also considered to be a good probe to study the viscous hydrodynamics behavior of the colliding system~\cite{v3hydro}.

Lots of properties of the triangular flow $v_3$ have been studied in hydrodynamic and transport models~\cite{v3hydro,v3ampt}. However, since the reaction plane can not be directly measured in the experiment, those anisotropic parameters can not be directly obtained. It is found that different methods may cause up to $20\%$  discrepancy on $v_2$~\cite{Artv2Review}, thus it should also be carefully evaluated for the $v_3$ study. Besides, $v_3$ is directly related with the initial fluctuation, it is interesting to see its system size dependence.

In this paper, we will study the triangular flow $v_3$ in both Au+Au and Cu+Cu collisions in a multiphase transport model (AMPT)~\cite{AMPT}. The relation between $v_3$ and $\varepsilon_{3}$ is studied as a function of number of participants and transverse momentum.  The paper is organized as follows: In Sec.~\ref{sect_obs}, the observables and technical methods are introduced. A brief description of AMPT model is given in Sec.~\ref{ampt_model}. The results and discussions are presented in Sec.~\ref{sect_discuss}. Finally, a summary is given in Sec.~\ref{sect_summary}.

\section{OBSERVABLES}
\label{sect_obs}

In a non-central collisions, the overlap region of two nuclei is an almond shape. Since the position of nucleons may fluctuate event by event, as discussed in Ref~\cite{epart2,definition, partecc}, those initial geometric irregularities of the colliding system can be described by $\varepsilon_{n}$:
\begin{equation}
\varepsilon _n  = \frac{{\sqrt {\left\langle {r^2 \cos (n\varphi
)} \right\rangle ^2 + \left\langle {r^2 \sin (n\varphi)}
\right\rangle ^2 } }}{{\left\langle {r^2 } \right\rangle }},
\label{eps}
\end{equation}
where $r$ and $\varphi$ are the polar coordinate position of participating nucleons and $\langle \cdots\rangle$ is the average over all the participants in an event. $n$ refers to the $n$-th harmonic, i.e., $\varepsilon_{2}$ describes the elliptic shape and $\varepsilon_{3}$ describes the triangular shape.


As the system evolves, the anisotropy in the coordinate space is transferred to the anisotropy in the momentum space due to the pressure gradient. The particle distribution then can be written as
\begin{equation}
\frac{dN}{d\phi}
\propto1+2\sum_{n=1}v_n\cos[n(\phi-\Psi_n)],
\label{eq1}
\end{equation}
where $\phi$ is the azimuthal angle, and $\Psi_{n}$ is the $n$-th event plane angle reconstructed by the final state particles:\emph{}
\begin{equation}
\Psi_n  = \frac{1}{n}\left[ \tan^{-1}\frac{\sum\limits_{i}\sin (n\phi_{i})}{\sum\limits_{i}\cos (n\phi_{i})}\right].
\label{eq3}
\end{equation}

The observed anisotropic flow is defined as the $n$-th Flourier coefficient $v_n$:
\begin{equation}
v^{\rm obs}_n=\left\langle\cos[n(\phi-\Psi_n)] \right\rangle.
\label{eq2}
\end{equation}
Here $\langle \cdots\rangle$ is taking the average over all the particles in the sample.

This is the so-called event plane method of calculating $v_n$. The reconstructed event plane fluctuates around the reaction plane. The observed signals need to be revised by the corresponding resolution~\cite{ArtPRC}:
\begin{equation}
v_n=\frac{v^{\rm obs}_n}{\mathscr{R}_n}.
\label{v2}
\end{equation}
Due to the finite multiplicity of final state particles, the resolution
\begin{equation}
\mathscr{R}_n = \langle \cos [n(\Psi_n-\Psi_{\rm nR})]\rangle
\label{res}
\end{equation}
is usually smaller than 1. $\Psi_{\rm nR}$ represents the nth real event plane angle.

\section{AMPT Model}
\label{ampt_model}
There are four main components in AMPT model:
the initial conditions, parton interactions, hadronization and hadron interactions. The initial conditions are obtained from the  HIJING model~\cite{HIJING},
which includes the spatial and momentum distributions of minijet partons from hard processes and
strings from soft processes. The time evolution of partons is then treated
according to the ZPC~\cite{ZPC} parton cascade model. After partons stop interacting,
a combined coalescence and string fragmentation model are used for the hadronization of partons.
The scattering among the resulting hadrons is described by a relativistic transport (ART) model~\cite{ART}
which includes baryon-baryon, baryon-meson and meson-meson elastic and inelastic scattering.

In our study, we analyzed the events from AMPT with the parton cross section equals to 3 mb and 10 mb. As all the
conclusions are independent on the parton cross section, only the results from 3 mb are shown in this paper.
There are about 8 million events in \auau collisions and 19 million events in \cucu collisions at \sNN = 200 used.
The string melting AMPT version is used since the previous study shows that the string melting AMPT version agrees with the experimental results better~\cite{AMPT}. The centrality is defined by the impact parameter.

\section{Results and Discussions}
\label{sect_discuss}

\begin{figure}[t]
\vskip 0cm
\includegraphics[width=0.5\textwidth]{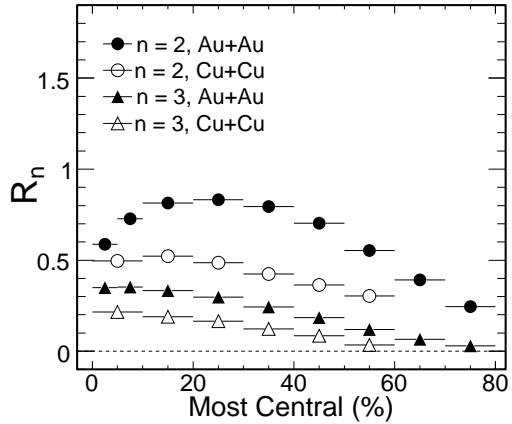}
\caption{The second and third harmonic event plane resolution calculated by the particles
 with pseudo-rapidty region of $|\eta| > 2$ as a function of centrality in both Au+Au and Cu+Cu collisions at \sNN=200 GeV in AMPT model.} \label{Plot::res}
\end{figure}

\begin{figure}[t]
\vskip 0cm
\includegraphics[width=0.5\textwidth]{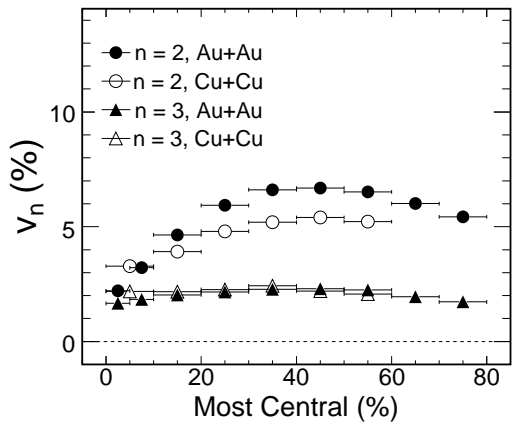}
\caption{$v_n$ as a function of centrality in both Au+Au and Cu+Cu collisions at \sNN=200 GeV in AMPT model.} \label{Plot::vn}
\end{figure}

In order to be comparable with the experimental data, the event plane method is used to calculate $v_n$. The procedure is slightly different between ours and Ref.~\cite{v3ampt}, in which the event plane was reconstructed by initial partons. Charged particles with $p_T\leq2$ GeV/$c$, $|\eta|>2$ are chosen to reconstruct the event plane according to the Eq.~\ref{eq3}.
The particles used for the $v_{n}$ measurements are within the $|\eta|<1$. The $\eta$ gap used here is to reduce the auto-correlation between the particles used to reconstruct the event plane and the particles of interest. In the following, the observed $v_{n}$ are all corrected by the corresponding resolution. Fig.~\ref{Plot::res} shows the resolution of $v_2$ and $v_3$ in both Au+Au and Cu+Cu collisions. The resolution of $v_2$ shows a peak in mid-central collisions which is consistent with the experimental result~\cite{PRCrun2}. This is because the resolution of $v_2$ is affected by both of the $v_2$ signal and the multiplicity used to reconstruct the event plane. While the resolution of $v_3$ only depends on the multiplicity, and keeps decreasing as the multiplicity drops.

In Fig.~\ref{Plot::vn}, $v_2$ and $v_3$ are shown as functions of centrality in both Au+Au and Cu+Cu collisions. We can see that $v_2$ shows strong centrality dependence since it is mainly coming from the elliptic anisotropy in the initial geometry. Unlike $v_2$, the dependence of $v_3$ on centrality and system size are much smaller. The trend of $v_3$ observed is the same as that in Ref~\cite{v3ampt}. However, the event plane angle $\Psi_n$ in Ref~\cite{v3ampt} is obtained from the initial parton distribution, which is not observed in the experiment, while in our study it is from the final state particle distribution. The results indicate that the triangular flow is less sensitive to the centrality and system
size compared with the elliptic flow. It could be understood as a result of combined effects from initial geometrical fluctuation and collective dynamics which requires the size of bulk to interact among themselves.

\begin{figure}
\vskip 0cm
\includegraphics[width=0.42\textwidth]{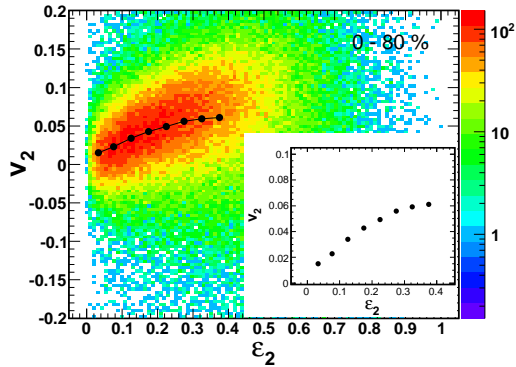}
\caption{(Color online) $v_2$ as a function of $\varepsilon_2$ in Au+Au collisions at \sNN=200 GeV in AMPT model. The black points are the average $v_2$ in the selected $\varepsilon_2$ bin. The pad in the right down corner is the average of $v_2$ with smaller scale.} \label{Plot::v2e2}
\end{figure}

\begin{figure}
\vskip 0cm
\includegraphics[width=0.42\textwidth]{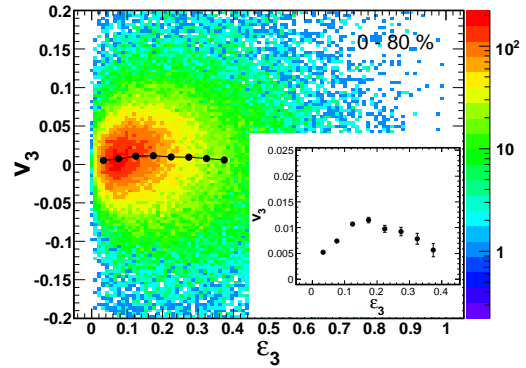}
\caption{(Color online)  $v_3$ as a function of $\varepsilon_3$ in Au+Au collisions at \sNN=200 GeV in AMPT model. The black points are the average $v_3$ in the selected $\varepsilon_3$ bin. The pad in the right down corner is the average of $v_3$ with smaller scale.} \label{Plot::v3e3}
\end{figure}

It is commonly assumed that the harmonic flow coefficients $v_n$ linearly depends on the $\varepsilon_n$. This
assumption is supported by hydrodynamic simulations~\cite{v3hydro} as long as one probes deformed initial profiles
with only a single non-vanishing harmonic eccentricity coefficient. In Fig.~\ref{Plot::v2e2} and Fig.~\ref{Plot::v3e3}, we investigate the feasibility of this assumption for $v_2$ and $v_3$ respectively. The relations between $v_n$ and $\varepsilon_n$ are drawn event by event in the two-dimension plots. The black points are the average values of $v_n$ in an selected $\varepsilon_n$ bin, and the curves are the connection of points to guide our eyes. The pads in the right down corners are the average values of $v_n$ with smaller scale. In Fig.~\ref{Plot::v2e2}, the $v_2$ increases with $\varepsilon_2$, which is consistent with the ideal hydrodynamic calculation~\cite{Ulrichv3hydro}. While in Fig.~\ref{Plot::v3e3}, the triangular flow $v_3$ firstly increases with $\varepsilon_3$ up to 0.17, and then decreases. Based on our study, the higher $\varepsilon_{3}$ bin corresponds to the more
peripheral collisions. It is known that $v_3$ is caused by initial geometrical fluctuation, and built up by the interactions of constituents. The less interactions in the higher $\varepsilon_{3}$ bin may cause the less converting efficiency from $\varepsilon_3$ to $v_3$.
That could be the reason of decreasing trend of $v_3$ when $\varepsilon_3$ is larger than 0.17. Both the trend and the value of $v_3$ show discrepancy to the ideal hydrodynamic calculations~\cite{Ulrichv3hydro}. As discussed in Ref.~\cite{v3hydro}, the viscosity causes the decrease of $v_3$, however, the effects to $v_3$ versus $\varepsilon_3$ is not
shown in the viscous hydrodynamic calculation.

\begin{figure}[t]
\vskip 0cm
\includegraphics[width=0.5\textwidth]{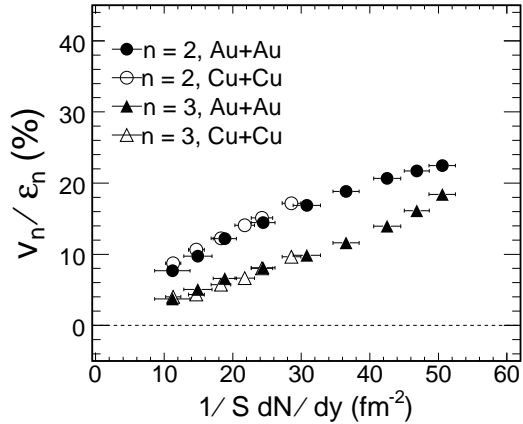}
\caption{$v_n/\varepsilon_n$ as a function of transverse particle density in both Au+Au and Cu+Cu collisions at \sNN=200 GeV in AMPT model.} \label{Plot::vnen}
\end{figure}

\begin{figure*}[ht]
\vskip 0cm
\includegraphics[width=0.8\textwidth]{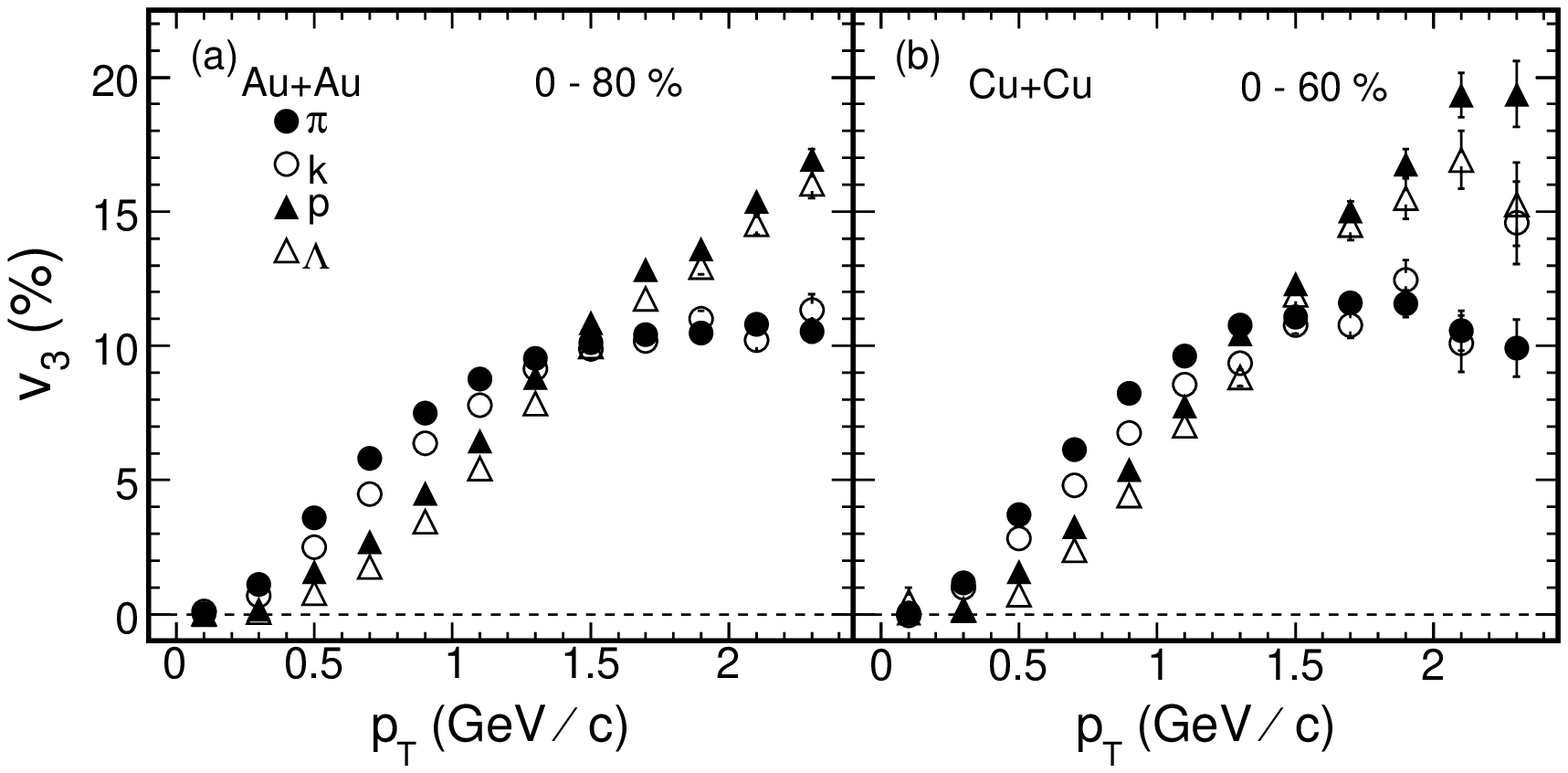}
\caption{$v_3$ as a function of transverse momentum in (a) Au+Au and (b) Cu+Cu collisions at \sNN=200 GeV in AMPT model.} \label{Plot::v3pt}
\end{figure*}

The ratio of elliptic flow to eccentricity $v_2/\varepsilon_2$ gains lots of interests by comparing with the hydrodynamic model~\cite{definition, Ollitrault_etaOverS}. Recently, the behavior of triangular flow $v_3$ in ideal hydrodynamics is also discussed~\cite{v3hydro}. In Fig.~\ref{Plot::vnen}, we study the $v_n/\varepsilon_n$ as a function of transverse particle density. From the plot we can see that $v_n/\varepsilon_n$ from Au+Au and Cu+Cu are consistent with each other very well. As the transverse particle density increases, $v_3/\varepsilon_3$ rises with smaller value than $v_2/\varepsilon_2$. It implies that as the particle density increases, the initial geometry asymmetry transfers to momentum asymmetry more efficiently while the system expands. Besides, the second order harmonic is more efficient than the third order.

At last, the transverse momentum dependence of $v_3$ for $\pi$, $K$, p and $\Lambda$ is also studied in Au+Au and Cu+Cu collisions. In Fig.~\ref{Plot::v3pt}, we can see that $v_3$ shows quite similar trend to $v_2$. At low $p_T$, the mass ordering phenomena is observed. The lighter particles are found with larger $v_3$.  It indicates that although $v_3$ is driven by $\varepsilon_3$, its transverse momentum dependence is dominated by the hydrodynamics behavior of the system. While when $p_T\geq1.5$ GeV/$c$, baryons and mesons are separated into two groups. The $p_T$ dependence of $v_3$ from identified particles is qualitatively similar to the $p_T$ dependence of $v_2$~\cite{PRCrun4, cucu_STAR}.
The $v_3$ results of identified particles from AMPT model are similar to the STAR preliminary results~\cite{Yadav}.

\section{Summary}
\label{sect_summary}

In summary, we studied the relation between initial geometry parameter $\varepsilon_n$ and anisotropic flow $v_n$ in Au+Au and
Cu+Cu collisions using the AMPT Monte-Carlo model. We find that the triangular flow $v_3$ is less sensitive to the centrality and system
size compared with the elliptic flow $v_2$.
The $v_2$ displays an increasing trend as a function of $\varepsilon_2$,
which is qualitatively consistent with hydrodynamic calculation. We found that $v_3$ shows an increasing trend when $\varepsilon_3$ is less than 0.17, and then
decreases beyond $\varepsilon_3$ = 0.17. It may be because of the lower converting efficiency from $\varepsilon_{3}$ to $v_{3}$ in the higher $\varepsilon_{3}$ bin.
This decreasing trend is in contrast to the results of ideal hydrodynamic calculation.
Both $v_2/\varepsilon_2$ and $v_3/\varepsilon_3$ increase with the transverse particle density,
and the second harmonic asymmetry in the initial geometry seems to transfer to the momentum asymmetry more efficiently than the third harmonic. The triangular flow $v_3$ of identified particles shows a mass ordering in low $p_T$ and meson-baryon splitting at intermediate $p_T$ in both Au+Au and Cu+Cu collisions which is similar to the $p_T$ dependence of $v_2$.

\vspace{7mm}

\section{Acknowledgments}

We wish to thank Prof. Fuqiang Wang for useful suggestions, and Dr. Kejun Wu for useful discussions on the AMPT model. This work was supported in part by the National Natural Science Foundation of China under grant No. 10775060, 11105060, 11135011, 11147146 and `the Fundamental Research Funds for the Central Universities', Grant No. HUST: 2011QN195.

%

%
\end{document}